\documentclass[letterpaper]{article} 
\usepackage{aaai2027}  
\usepackage[hyphens]{url}  
\usepackage{graphicx} 
\usepackage{natbib}  
\usepackage{caption} 
\usepackage{algorithm}
\usepackage{algorithmic}

\usepackage{newfloat}
\usepackage{listings}
\DeclareCaptionStyle{ruled}{labelfont=normalfont,labelsep=colon,strut=off} 
\floatstyle{ruled}
\newfloat{listing}{tb}{lst}{}
\floatname{listing}{Listing}

\usepackage{booktabs}
\usepackage{amsmath}
\usepackage{amssymb}
\usepackage{amsfonts}
\usepackage{multirow}
\usepackage{xcolor}
\title{CARA: Cognitive Adaptive Recommendation Agent}

\author{
    Weijun Gao\textsuperscript{\rm 1}\equalcontrib,
    Jinyang Dong\textsuperscript{\rm 2}\equalcontrib,
    Chuanru Ren\textsuperscript{\rm 2},
    Hengxiao Li\textsuperscript{\rm 2}
}

\affiliations{
    \textsuperscript{\rm 1}The Chinese University of Hong Kong\\
    \textsuperscript{\rm 2}Tongji University
}

\begin{document}

\maketitle

\begin{abstract}
Recent advances in large language models and agent-based recommendation frameworks have introduced new opportunities for more flexible and context-aware recommendation. However, existing methods still largely rely on semantic matching, end-to-end generation, or loosely structured agent workflows, without explicitly modeling how user preferences are processed and translated into final decisions. To address this limitation, we propose CARA, a cognitively inspired recommendation framework that formulates recommendation as a structured decision-making process. The core intuition of CARA is that user decisions are jointly shaped by two complementary mechanisms: intuitive affective preference and deliberate rational evaluation. Accordingly, CARA organizes recommendation into two coordinated stages: candidate filtering, which narrows the search space based on coarse-grained preference constraints, and dual-perspective decision modeling, which captures recommendation decisions through affective and rational judgment. We further introduce a boundary-aware KTO strategy that prioritizes instructions the model can solve occasionally but not consistently, thereby increasing the density of informative preference signals. Extensive experiments on three Amazon Reviews domains show that CARA achieves the best performance on most evaluation metrics, with relative improvements of up to 10.15\% over the baseline.
\end{abstract}
\section{Introduction}

Recommender systems have become an important component of modern information ecosystems. They are widely deployed in e-commerce, content platforms, and social media services to help users discover items of interest from large-scale item collections.The development of large language models provides a new technical route for recommender systems ~\cite{zhao2023llmsurvey,zhao2024recsysllm,lin2025benefit}. Owing to their strong language understanding, knowledge transfer, and reasoning capabilities, LLMs can jointly process user histories, product texts, and other information. However, existing LLM-based recommendation agents ~\cite{cheng2024exploring,wang2024surveyagents,xi2025rise} still face several limitations. Many methods primarily rely on semantic matching and lack explicit modeling of the recommendation decision process. They usually feed user profiles, historical behaviors, and candidate item information into a single reasoning chain and directly generate the ranking result, which can easily mix decision evidence of different natures.

Real user choices are jointly shaped by candidate feasibility judgment, multi-dimensional preference evaluation, and trade-offs among different pieces of evidence. To model this process, we propose CARA (\textbf{C}ognitive \textbf{A}daptive \textbf{R}ecommendation \textbf{A}gent), an adaptive recommendation agent framework inspired by cognitive decision-making processes ~\cite{kahneman2011thinking} . This view distinguishes between fast and intuitive preference formation and slower, more deliberative evaluation. CARA operationalizes these complementary mechanisms as affective and rational judgment within a coarse-to-fine recommendation process.

Specifically, the system first performs grounded candidate filtering based on user memory and candidate information. It then evaluates the retained candidates from two complementary perspectives. Affective judgment captures users' preferences for categories, features, styles, and usage scenarios, whereas rational judgment evaluates price reasonableness, budget compatibility, quality reliability, and practical utility.

To improve the agent's task accuracy, output stability, and factual
consistency, we post-train CARA through supervised fine-tuning (SFT)
followed by boundary-aware Kahneman--Tversky Optimization
(KTO). SFT teaches the
agent basic recommendation logic and structured output formats. For
KTO, we sample multiple responses for each instruction and estimate its
empirical solvability according to format validity, decision correctness,
and factual consistency. Instructions that are solved consistently
provide limited additional learning signals, whereas instructions that
are rarely solved tend to yield noisy feedback beyond the model's
current capability. We therefore retain instructions with intermediate
empirical solvability, which the model can solve occasionally but has
not yet mastered consistently. By concentrating preference optimization
on these boundary samples, the proposed strategy increases the density
of informative training signals and improves decision reliability and
factual grounding.

We conduct experiments on three domains from Amazon Reviews dataset: CDs, Office, and Beauty. The results show that CARA achieves strong performance across all three domains and outperforms representative baselines on most evaluation metrics. Ablation studies demonstrate the effectiveness of candidate filtering, affective judgment, and rational judgment. Training-stage analysis shows that SFT improves basic task capabilities, while boundary-aware KTO further improves judgment stability on complex samples and reduces hallucination rates.

Our main contributions are summarized as follows:

\begin{itemize}
\item We propose CARA, a cognitively adaptive recommendation framework that formulates recommendation as a coarse-to-fine structured decision process. CARA integrates grounded candidate filtering, affective and rational judgment, confidence-adaptive fusion, and error-driven updates of dual cognitive memories.

\item We propose a boundary-aware preference optimization strategy that selects instructions with intermediate empirical solvability for KTO. By focusing optimization on informative samples near the model's current capability boundary, the proposed strategy improves task accuracy, output stability, and factual consistency.

\item Extensive experiments on three Amazon Reviews domains show that CARA ranks first or second in all 12 evaluation settings and achieves relative gains of up to 10.15\% over the strongest competing baselines. Ablation and post-training analyses further validate the effectiveness of its core components.

\end{itemize}

\section{Related Work}

\subsection{Classical Recommendation Systems}
Recommender systems have long relied on user-item interactions to learn personalized preferences. Early collaborative filtering and matrix factorization methods perform personalized ranking by modeling latent representations of users and items~\cite{rendle2009bpr,koren2009matrix}, while simple methods such as Pop and BM25~\cite{cremonesi2010performance,robertson2009probabilistic} use item popularity and textual relevance as recommendation signals, respectively. With the development of deep learning~\cite{he2017ncf}, sequential recommendation models further regard user history as a temporally evolving behavior sequence. GRU4Rec~\cite{hidasi2016gru4rec} captures short-term interest evolution through recurrent neural networks, while self-attention-based models capture long-range dependencies in the sequence~\cite{kang2018sasrec}. Recent studies further improve sequential recommendation by scaling sequence modeling architectures and capturing latent user intentions, HSTU reformulates large-scale recommendation as a sequential transduction task~\cite{zhai2024actions}, while intent contrastive learning enhances user preference modeling through cross-subsequence representations~\cite{qin2024intent}.

\subsection{LLM-Based Agents}
LLM-based agents have become an important research direction due to their ability to interact with environments through natural language, use tools, maintain memory, and perform multi-step reasoning~\cite{cheng2024exploring,wang2024surveyagents,xi2025rise}. Existing studies enhance agent capabilities through reasoning-action synergy, personalized memory, and experiential feedback~\cite{yao2023react,park2023generativeagents,zhao2024expel}. 

Beyond individual agents, multi-agent frameworks further improve problem solving through role specialization, collaborative planning, and conversational coordination~\cite{chen2024agentverse,hong2024metagpt,wu2023autogen}. In parallel, deliberate reasoning methods such as chain-of-thought, tree-structured reasoning, and graph-structured reasoning demonstrate that explicit intermediate reasoning can improve LLMs' ability to solve complex tasks~\cite{wei2022cot,yao2023tree,besta2024graph}.

\begin{figure*}[t]
    \centering
    \includegraphics[width=\textwidth]{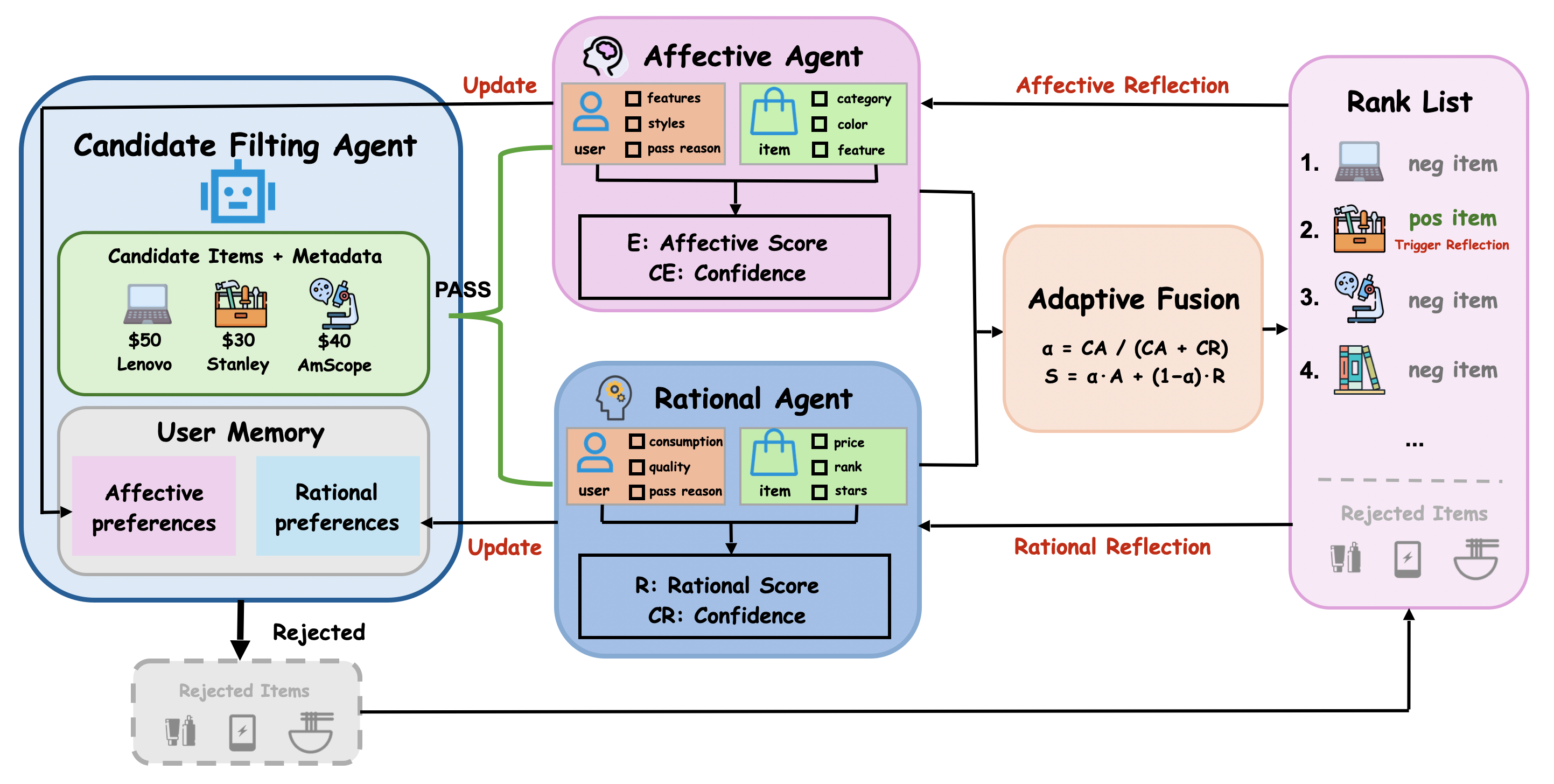}
    \caption{Overall architecture of the proposed CARA.}
    \label{fig:cara_framework}
\end{figure*}

\subsection{LLM for Recommender Systems}

The application of LLMs to recommender systems represents a profound shift in research paradigm, mainly following two technical routes: constructing direct end-to-end LLM recommendation frameworks and using LLMs to enhance traditional recommendation models ~\cite{cui2022m6rec,geng2022p5,zhao2023llmsurvey}. These methods expand the technical boundaries of recommendation and provide new solutions to long-standing problems such as data sparsity and cold start. End-to-end systems aim to transform recommendation tasks into language modeling problems and unify the recommendation pipeline ~\cite{geng2024breaking}. Methods such as LLMRank further use advanced prompting to achieve zero-shot or few-shot recommendation, enabling LLMs to directly rank candidate items ~\cite{hou2024llmrank}. Meanwhile, enhancement-based methods focus on integrating the open-world knowledge of LLMs with collaborative signals, or using LLM reasoning abilities to understand user intentions and map preferences ~\cite{sun2024llmcf,xi2024openworld,yang2024itemlm}. RecMind and AgentCF simulate recommendation through agent-style reasoning and collaboration ~\cite{wang2024recmind,zhang2024agentcf}, while STARec introduces autonomous deliberate reasoning for efficient LLM-agent recommendation~\cite{wu2025starec}.PromptRec reformulates cold-start recommendation through prompt-based language modeling ~\cite{wu2024promptrec}, and TaxRec leverages taxonomy and attribute information to enhance the matching between user preferences and item representations ~\cite{liang2025taxrec}. Although these methods significantly improve recommender systems' ability to utilize textual information and open semantic knowledge, most existing work still relies on holistic prompting, semantic matching, or end-to-end ranking, and remains insufficient in structurally decomposing different preference factors in user decision-making.

\section{Methodology}

\subsection{Problem Definition}
We formulate recommendation as an agent-based candidate ranking task. Let $\mathcal{U}$ denote the user set, $\mathcal{I}$ denote the item set, and $\mathcal{X}$ denote the available item information. For a user $u\in\mathcal{U}$, the temporally ordered interaction history is denoted as $\mathcal{H}_u^t=\{i_u^1,i_u^2,\ldots,i_u^{t-1}\}$. Given the user history, the candidate set $\mathcal{C}_u^t\subseteq\mathcal{I}$, and the available information of candidate items, CARA uses multiple collaborative agents to analyze user preferences and generate a ranking list $\hat{\mathcal{R}}_u^t$. The goal is to rank the ground-truth item $i_u^{+}$ as highly as possible in $\hat{\mathcal{R}}_u^t$.

\subsection{Structure Overview}

CARA is inspired by the dual-process view of human decision-making, which distinguishes intuitive preference formation from more deliberative evaluation. CARA operationalizes these complementary processes as affective judgment and rational judgment within a coarse-to-fine recommendation architecture.

Specifically, CARA consists of grounded candidate filtering, dual-perspective decision-making, confidence-adaptive fusion, and feedback-driven memory update. The candidate filtering agent first removes items that clearly violate the user's needs or constraints. The retained candidates are then evaluated through affective judgment, which captures intuitive preference matching, and rational judgment, which considers explicit constraints and utility. Their outputs are adaptively fused according to judgment confidence. When the resulting recommendation is inconsistent with the observed user behavior, CARA attributes the error and updates the corresponding affective or rational memory.

\subsubsection{Memory Architecture}

To provide each agent with continuous and personalized decision evidence, CARA equips each user with a cognitive memory module represented in natural language. This module stores the agent's latest summary of user preferences retrieved from historical behaviors. The cognitive memory of user $u$ at time step $t$ is represented as $\mathcal{M}_u^t=(\mathcal{M}_{u,A}^t,\mathcal{M}_{u,R}^t)$, where $\mathcal{M}_{u,A}^t$ and $\mathcal{M}_{u,R}^t$ denote affective memory and rational memory.

Affective memory mainly describes the user's preferences for product categories, features, and styles. Rational memory mainly describes the user's acceptable price range and quality requirements, providing evidence for budget compatibility and item reliability judgments.

\subsubsection{Candidate Filtering}

Directly performing complex cognitive judgment over all candidate items introduces additional reasoning cost and may allow clearly irrelevant items to interfere with the final decision. Therefore, CARA first introduces a candidate filtering agent to perform coarse-grained filtering over the candidate set. For a candidate item $i$, this agent receives the user's combined cognitive memory $\mathcal{M}_u^t$ together with the item's textual and statistical information, and outputs a filtering decision $d_i\in\{\mathrm{PASS},\mathrm{REJECT}\}$ and a corresponding reason $r_i$.

The candidate filtering agent performs conservative screening based on explicit and high-confidence constraints, such as clear budget violations and evident conflicts with established user preferences. Only candidates with sufficient evidence of infeasibility are assigned \texttt{REJECT}, while ambiguous candidates are retained for further evaluation. Candidates assigned \texttt{PASS}, together with a concise screening rationale $r_i$, are passed to the affective and rational judgment modules. This stage narrows the candidate set before fine-grained evaluation and provides both modules with a shared evidence base, reducing redundant downstream reasoning.

\subsubsection{Dual-Perspective Decision Modeling}
Following the dual-process view of human decision-making, CARA decomposes recommendation decisions into two complementary perspectives: affective preference and rational utility. 
Specifically, affective judgment focuses on subjective and context-sensitive attributes that reflect personal preference, whereas rational judgment focuses on objective and constraint-related attributes that determine feasibility and practical utility. For each candidate item $i$, the two judgment modules receive the corresponding cognitive memory, item information, and the reason $r_i$ provided by the candidate filtering agent, and output a judgment score and confidence:
\begin{equation}
    (s_{i,k},c_{i,k})
    =
    f_k
    \left(
    \mathcal{M}_{u,k}^t,
    \mathbf{x}_i^k,
    r_i
    \right),
    \quad
    k\in\{A,R\}.
\end{equation}
Here, $A$ and $R$ denote the affective and rational perspectives. $\mathbf{x}_i^A$ denotes the item textual information used for affective judgment, while $\mathbf{x}_i^R$ denotes the item statistical information used for rational judgment. $s_{i,k}$ and $c_{i,k}$ denote the judgment score and confidence under perspective $k$.

Under different item and user states, the reliability of affective evidence and rational evidence may vary. Therefore, CARA dynamically computes the affective weight according to the confidence values of the two modules:
\begin{equation}
    \alpha_i =
    \frac{c_{i,A}}{c_{i,A}+c_{i,R}}.
\end{equation}
The final decision score of a candidate item is defined as
\begin{equation}
    S(u,i)
    =
    \alpha_i{s_{i,A}}
    +
    (1-\alpha_i){s_{i,R}}.
\end{equation}

\subsubsection{Reflection and Memory Update}

During sequential update, CARA performs error-driven cognitive reflection to analyze the cause of a current recommendation failure and determine which type of user memory should be revised. Given the ranking result at time step $t$, CARA compares the top-ranked predicted item $\hat{i}_u^t$ with the truly interacted target item $i_u^{+}$. A ranking error triggers reflection and a selective update of the corresponding memory.

The reflection process centers on error attribution and then selects the corresponding memory update path. If the target item is filtered out during candidate filtering, it indicates that the current coarse-grained preference constraints may be overly strict; the system then retrieves matching evidence between the target item and the user's history to revise the candidate filtering boundary. If the wrong item receives a higher score in affective judgment, CARA compares the differences between the target item and the wrongly recommended item in terms of category, feature, style, and usage scenario, and retrieves supporting evidence from the user's recent interactions to update affective memory. If the wrong item receives a higher score in rational judgment, CARA checks whether the current decision violates the user's price range, quality requirements, or other explicit constraints, and revises rational memory using historical consumption statistics and item quality evidence. In this way, CARA transforms recommendation errors into structured feedback for candidate filtering, affective preference, or rational judgment, rather than treating them as undifferentiated ranking failures.

Formally, let $\mathcal{G}_u^t$ denote the evidence used in the current reflection, including the target item, the wrongly recommended item, candidate filtering results, dual-perspective judgment rationales, and the retrieved user historical preference information. The memory update process is represented as
\begin{equation}
    \mathcal{M}_u^{t+1}
    =
    \operatorname{Reflect}
    \left(
    \mathcal{M}_u^t,
    i_u^{+},
    \hat{i}_u^t,
    \mathcal{G}_u^t
    \right),
\end{equation}
where $\hat{i}_u^t$ denotes the item that is incorrectly preferred in the current decision process. To avoid over-correction caused by a single feedback signal, CARA adopts a conservative update strategy: it only modifies erroneous memory fragments that conflict with true feedback or historical evidence, while preserving summarized long-term stable preferences. Thus, CARA can continuously revise the user's cognitive state along sequential interactions without updating the parameters of the underlying large model.In the offline evaluation of this work, this update mechanism is applied only during the historical memory-evolution phase, while user memories remain fixed for all test interactions.

\subsection{CARA Post-Training}

To enable CARA to make stable and reliable decisions in real recommendation pipelines, we further optimize the agent through supervised fine-tuning and reinforcement learning. Specifically, supervised fine-tuning is mainly used to establish basic task understanding, format stability, and initial judgment accuracy, allowing the agent to generate parseable recommendation decisions according to a predefined structure ~\cite{ouyang2022training}. The reinforcement learning stage further improves the model's factual consistency and task accuracy in complex candidate scenarios, reducing hallucinated judgments that lack support from historical evidence. Unlike directly performing preference optimization on all training samples, we propose a boundary-aware KTO training strategy that concentrates optimization on training samples that the model has not yet mastered stably but that remain learnable, thereby improving training efficiency and preference alignment quality ~\cite{ethayarajh2024kto}.

\subsubsection{Supervised Fine-Tuning}

The first step toward effective supervised fine-tuning (SFT) is to obtain high-quality training data. To avoid data leakage, we select a new set of users isolated from the main experiments to construct candidate samples. For each training sample, we use a stronger teacher model to generate a user profile based on the user history, and further generate a structured response containing the analysis process, recommendation rationale, and final decision. We then remove responses with format errors, hallucinations, or incomplete content through rule-based checks.

Given the SFT dataset $\mathcal{D}_{\mathrm{SFT}}=\{(q_n,y_n)\}_{n=1}^{N}$, where $q_n$ denotes the instruction composed of user history, item information, and task requirements, and $y_n$ denotes the structured response generated by the teacher model, the agent is optimized with the standard autoregressive objective:
\begin{equation}
    \mathcal{L}_{\mathrm{SFT}}
    =
    -\sum_{(q,y)\in\mathcal{D}_{\mathrm{SFT}}}
    \sum_{t=1}^{|y|}
    \log \pi_{\theta}(y_t\mid q,y_{<t}).
\end{equation}
This stage teaches the agent how to understand the recommendation context and follow the output format. In other words, SFT provides an initialization policy with basic task accuracy and format stability for preference optimization.

\subsubsection{Boundary-Aware KTO Optimization}

After SFT, we further use KTO to optimize the agent with preference signals~\cite{ethayarajh2024kto}. A direct approach is to follow the standard KTO training pipeline by constructing one desirable and one undesirable response for each instruction and optimizing over all instructions. However, in preliminary experiments, we find that this naive strategy does not consistently reduce the model's hallucination rate, and may even increase hallucination in some datasets. Further analysis shows that the problem mainly comes from the effective signal density of the preference optimization data. After SFT, the model can already generate format-correct, decision-correct, and factually consistent responses for many simple instructions. The negative feedback in these samples usually contains only minor errors and thus provides weak reverse learning signals. Meanwhile, a small number of extremely difficult samples often exceed the current capability of the model, and their negative feedback is also difficult to translate into effective policy improvement. Therefore, directly using all instructions dilutes truly valuable boundary samples with a large number of easy samples and noisy samples, thereby weakening the optimization effect of KTO.

Based on this observation, we propose a boundary-aware KTO strategy. The core idea is that preference optimization should not be applied uniformly to all instructions, but should prioritize decision-boundary regions where the model can produce correct answers but remains unstable. Specifically, for each instruction $q_n$, we sample $m$ independent responses $\{y_{n,j}\}_{j=1}^{m}$ from the SFT model and evaluate each response from three dimensions: format validity, task decision correctness, and factual consistency. Let $F_{n,j}$, $D_{n,j}$, and $H_{n,j}$ denote format validity, decision correctness, and hallucination judgment, respectively. The empirical solvability of an instruction is defined as
\begin{equation}
\begin{aligned}
    \hat{p}_n
    &=
    \frac{1}{m}
    \sum_{j=1}^{m}
    \mathbb{I}
    \left[
    F_{n,j}=1
    \land D_{n,j}=1
    \land H_{n,j}=0
    \right], \\
    \mathcal{Q}_{\mathrm{bd}}
    &=
    \left\{
    q_n \mid \tau_{\min} \leq \hat{p}_n \leq \tau_{\max}
    \right\}.
\end{aligned}
\end{equation}
Here, $\mathcal{Q}_{\mathrm{bd}}$ denotes the selected boundary instruction set. When $\hat{p}_n$ is too high, the model has already mastered the sample stably and further optimization brings limited benefit. When $\hat{p}_n$ is too low, the model can hardly generate reliable responses, and its feedback signal may be overly noisy. Therefore, we only retain instructions whose empirical solvability lies within $[\tau_{\min},\tau_{\max}]$, thereby increasing the proportion of effective learning signals in the preference optimization data.

For each retained instruction, we select two desirable responses as positive feedback samples and two undesirable responses as negative feedback samples. A desirable response must simultaneously satisfy format correctness, decision correctness, and the absence of obvious hallucination, while an undesirable response contains format errors, decision errors, or a lack of factual support. Finally, we obtain the KTO dataset $\mathcal{D}_{\mathrm{KTO}}$.

During optimization, we use the SFT model as the reference policy $\pi_{\mathrm{ref}}$ and update the current policy $\pi_{\theta}$ with the KTO objective. Unlike DPO~\cite{rafailov2023dpo}, which relies on pairwise preference comparison, KTO maps desirable and undesirable samples into asymmetric utilities, thereby directly exploiting binary feedback signals for optimization. Let
$r_{\theta}(q,y)=\beta\log
\frac{\pi_{\theta}(y\mid q)}
{\pi_{\mathrm{ref}}(y\mid q)}$.
The KTO objective function is defined as follows:
\begin{equation}
\begin{aligned}
J_{\mathrm{KTO}}(\theta)
=
&\mathbb{E}_{\mathcal{D}_{\mathrm{KTO}}^{+}}
\left[
\lambda_{+}\sigma
\left(r_{\theta}(q,y^{+})-z_{\mathrm{ref}}\right)
\right]
\\
&+
\mathbb{E}_{\mathcal{D}_{\mathrm{KTO}}^{-}}
\left[
\lambda_{-}\sigma
\left(z_{\mathrm{ref}}-r_{\theta}(q,y^{-})\right)
\right].
\end{aligned}
\end{equation}

Through this boundary-aware data construction strategy, KTO is no longer dominated by a large number of simple samples that the model has already mastered. Instead, it focuses on optimizing the model's unstable behaviors near the decision boundary. This training paradigm can more effectively reduce format drift, factual inconsistency, and wrong judgments, thereby improving the agent's output stability, grounding ability, and task accuracy.

\section{Experiments}

\subsection{Experimental Setup}

\subsubsection{Datasets}
We evaluate CARA on three Amazon Reviews domains~\citep{mcauley2015image}:
CDs, Office, and Beauty. For each domain, we chronologically organize
the interactions of 100 users, using the first 10 interactions for
cognitive memory evolution and the remaining interactions for testing.
Detailed statistics are provided in supplementary material.


\begin{table*}[!htbp]
    \centering
    \setlength{\tabcolsep}{0pt}
    \begin{tabular*}{\textwidth}{@{\extracolsep{\fill}}l*{12}{c}@{}}
        \toprule
        \multirow{2}{*}{Method}
        & \multicolumn{4}{c}{Office}
        & \multicolumn{4}{c}{CDs}
        & \multicolumn{4}{c}{Beauty} \\
        \cmidrule(lr){2-5}\cmidrule(lr){6-9}\cmidrule(lr){10-13}
        & HR@1 & HR@5 & N@5 & N@10
        & HR@1 & HR@5 & N@5 & N@10
        & HR@1 & HR@5 & N@5 & N@10 \\
        \midrule
        Pop$_{\mathrm{Full}}$
        & .0510 & .4497 & .2404 & .4155
        & .1325 & .4970 & .3145 & .4724
        & .0751 & .4585 & .2584 & .4303 \\
        BPR$_{\mathrm{Full}}$
        & .0952 & .5545 & .3183 & .4614
        & \underline{.1834} & .6225 & .4042 & .5259
        & .1592 & \textbf{.6917} & \underline{.4196} & .5201 \\
        BPR$_{\mathrm{Sample}}$
        & .1186 & .5379 & .3224 & .4686
        & .0746 & .4970 & .2811 & .4416
        & .1031 & .5045 & .3009 & .4596 \\
        BM25
        & .1628 & .5779 & .3701 & .5070
        & .1811 & .6059 & .3875 & .5178
        & \textbf{.1892} & .6396 & .4122 & \underline{.5284} \\
        \midrule
        GRU4Rec
        & .0841 & .4497 & .2609 & .4362
        & .0769 & .4213 & .2382 & .4160
        & .0701 & .4204 & .2366 & .4217 \\
        SASRec
        & .1090 & .4772 & .2888 & .4547
        & .1065 & .5041 & .2966 & .4554
        & .1211 & .5215 & .3156 & .4676 \\
        \midrule
        LLMRank
        & .1269 & .6041 & .3699 & .4966
        & .0994 & .5373 & .3190 & .4664
        & .1061 & .5826 & .3455 & .4800 \\
        AgentCF
        & \textbf{.2041} & .6097 & .4066 & .5329
        & .1124 & .5302 & .3184 & .4699
        & .1471 & .5896 & .3664 & .4980 \\
        PromptRec
        & .1241 & .5834 & .3501 & .4834
        & .1136 & .5479 & .3231 & .4678
        & .1101 & .5205 & .3101 & .4626 \\
        TaxRec
        & .2000 & \underline{.6662} & \underline{.4401} & \underline{.5484}
        & \underline{.1834} & \underline{.6308} & \underline{.4073} & \underline{.5269}
        & .1642 & .5996 & .3811 & .5092 \\
        \midrule
        \textbf{CARA}
        & \underline{.2028} & \textbf{.7338} & \textbf{.4782} & \textbf{.5646}
        & \textbf{.1894} & \textbf{.6734} & \textbf{.4341} & \textbf{.5378}
        & \underline{.1822} & \underline{.6547} & \textbf{.4201} & \textbf{.5297} \\
        \bottomrule
    \end{tabular*}
    \caption{Overall performance comparison. Bold numbers denote the best results and
    underlined numbers denote the second-best results among all methods.}
    \label{tab:main_results}
\end{table*}

\subsubsection{Evaluation Metrics}

For each test interaction, we construct a candidate set containing one
ground-truth item and nine items that the user has not interacted with.
Negative items are drawn from a pre-generated fixed candidate pool, so
all methods are evaluated on the same candidates. We evaluate ranking
performance using Hit Ratio (HR) and Normalized Discounted Cumulative
Gain (NDCG). Since each candidate set contains only one positive item,
NDCG@1 is equivalent to HR@1; therefore, we report HR@1, HR@5,
NDCG@5, and NDCG@10. We additionally report the hallucination rate,
defined as the proportion of generated responses containing at least
one factual claim about user preferences, interaction history, or
candidate attributes that is unsupported by or inconsistent with the
input. Recommendation and format errors are evaluated separately.

\subsubsection{Baseline Methods}

We compare CARA with representative baselines from three categories: traditional recommendation and retrieval methods, including Pop, BPR, and BM25; sequential recommendation models, including GRU4Rec, SASRec; and LLM-based recommendation methods, including LLMRank~\cite{hou2024llmrank}, AgentCF~\cite{zhang2024agentcf}, PromptRec~\cite{wu2024promptrec}, and TaxRec~\cite{liang2025taxrec}. The suffixes \textit{Full} and \textit{Sample} denote training on all
users and on the same 100-user subset as CARA.

\subsubsection{Implementation Details}

CARA uses Qwen3-1.7B as the recommendation agent and Qwen3-32B as
both the teacher and judge models. We set the KTO
boundary interval to $[0.35,0.65]$, select two desirable and two
undesirable responses for each retained instruction. Additional implementation details
are provided in the supplementary material.

\subsection{Overall Performance}

Table~\ref{tab:main_results} compares CARA with representative baselines. Overall, CARA achieves consistently competitive performance across the three domains, with particularly clear advantages on metrics that evaluate the quality of the entire ranking list. These results highlight the benefit of explicitly modeling user decisions under sparse interaction settings.

Classical sequential recommendation models, including GRU4Rec and SASRec, do not exhibit stable advantages in our setting. Although these methods are designed to capture sequential interaction patterns, the limited history length and small user population under the \textit{Sample} setting provide insufficient collaborative signals for learning reliable user-item representations. CARA is less dependent on such signals, as it performs structured reasoning over user memory, item descriptions, and statistical attributes. This enables it to adapt more effectively to few-user and near-cold-start scenarios.

Methods that exploit textual semantics, such as BM25, AgentCF, and TaxRec, are generally more competitive than purely interaction-based sequential models. Their strong performance on several metrics confirms that item descriptions and language-model knowledge can partially alleviate data sparsity. Nevertheless, these approaches typically perform semantic matching or holistic ranking without explicitly distinguishing different factors underlying user decisions. CARA instead decomposes the process into candidate filtering, affective judgment, and rational judgment, providing more structured evidence for ranking.

This advantage is most evident in the overall ranking quality rather than top-1 accuracy alone. On Office and CDs, CARA achieves larger gains at deeper cutoffs, suggesting that it more consistently moves the target item toward the top of the candidate list. On Beauty, BM25 and BPR$_{\mathrm{Full}}$ remain competitive on several top-rank metrics, indicating that lexical matching and collaborative signals can capture some dominant preference patterns in this domain. CARA performs better on NDCG@5 and NDCG@10, demonstrating a more reliable ordering of the candidate set as a whole.

\subsection{Model analysis}

\begin{table*}[!htbp]
    \centering
    \setlength{\tabcolsep}{0pt}
    \begin{tabular*}{\textwidth}{@{\extracolsep{\fill}}l*{12}{c}@{}}
        \toprule
        \multirow{2}{*}{Variant}
        & \multicolumn{4}{c}{Office}
        & \multicolumn{4}{c}{CDs}
        & \multicolumn{4}{c}{Beauty} \\
        \cmidrule(lr){2-5}\cmidrule(lr){6-9}\cmidrule(lr){10-13}
        & HR@1 & HR@5 & N@5 & N@10
        & HR@1 & HR@5 & N@5 & N@10
        & HR@1 & HR@5 & N@5 & N@10 \\
        \midrule
        CARA
        & \textbf{.2028} & \textbf{.7338} & \textbf{.4782} & \textbf{.5646}
        & \textbf{.1894} & \textbf{.6734} & \textbf{.4341} & \textbf{.5378}
        & \textbf{.1822} & \textbf{.6547} & \textbf{.4201} & \textbf{.5297} \\
        w/o Filtering
        & .1738 & .6497 & .4076 & .5205
        & .1645 & .5870 & .3747 & .5073
        & .1461 & .6166 & .3782 & .5025 \\
        w/o Rational
        & .1301 & .5604 & .3404 & .4823
        & .1302 & .5692 & .3410 & .4786
        & .1361 & .5446 & .3341 & .4795 \\
        w/o Affective
        & .1214 & .5586 & .3358 & .4786
        & .1254 & .5183 & .3195 & .4720
        & .1001 & .5125 & .3038 & .4584 \\
        \bottomrule
    \end{tabular*}
    \caption{Ablation results of CARA on three datasets.}
    \label{tab:ablation_results}
\end{table*}

\begin{table*}[t]
    \centering
    \setlength{\tabcolsep}{0pt}
    \begin{tabular*}{\textwidth}{@{\extracolsep{\fill}}l*{12}{c}@{}}
        \toprule
        \multirow{2}{*}{Model}
        & \multicolumn{4}{c}{Office}
        & \multicolumn{4}{c}{CDs}
        & \multicolumn{4}{c}{Beauty} \\
        \cmidrule(lr){2-5}\cmidrule(lr){6-9}\cmidrule(lr){10-13}
        & HR@1 & HR@5 & N@5 & N@10
        & HR@1 & HR@5 & N@5 & N@10
        & HR@1 & HR@5 & N@5 & N@10 \\
        \midrule
        Base
        & .1751 & .6069 & .3910 & .5169
        & .1645 & .5823 & .3706 & .5041
        & .1551 & .6076 & .3830 & .5082 \\
        SFT
        & \underline{.1986} & \underline{.6359} & \underline{.4221} & \underline{.5395}
        & \underline{.1831} & \underline{.6237} & \underline{.4020} & \underline{.5218}
        & \textbf{.1911} & \underline{.6106} & \underline{.4062} & \textbf{.5302} \\
        Ours
        & \textbf{.2028} & \textbf{.7338} & \textbf{.4782} & \textbf{.5646}
        & \textbf{.1894} & \textbf{.6734} & \textbf{.4341} & \textbf{.5378}
        & \underline{.1822} & \textbf{.6547} & \textbf{.4201} & \underline{.5297} \\
        \bottomrule
    \end{tabular*}
    \caption{Recommendation performance with different agent training stages.}
    \label{tab:alignment_results}
\end{table*}

\subsubsection{Ablation Study}

To analyze the independent contribution of each module, we construct three ablated variants: removing the candidate filtering module (w/o Filtering), removing rational judgment (w/o Rational), and removing affective judgment (w/o Affective). The results are shown in Table~\ref{tab:ablation_results}.

As can be observed, all three ablated variants perform worse than the full CARA on all datasets and metrics, indicating that candidate filtering, rational judgment, and affective judgment all contribute to final recommendation performance. Removing the candidate filtering module weakens the model's preliminary constraint over the candidate space. Removing rational or affective judgment breaks the dual-perspective decision structure and forces the model to rely on a single type of preference evidence. Overall, the results show that CARA's performance improvement comes from the collaboration of multiple cognitive modules rather than from an accidental gain brought by a single module.

\subsubsection{Effects of SFT and KTO}
\begin{figure}[t]
    \centering
    \includegraphics[width=0.95\columnwidth]{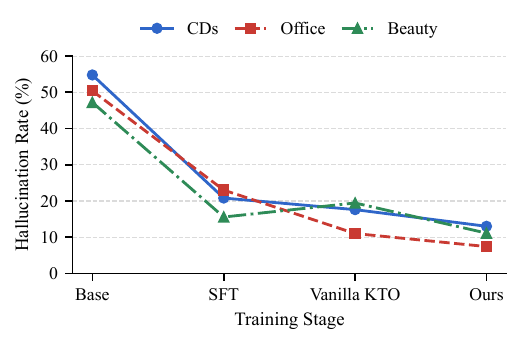}
    \caption{Hallucination rates across training stages. ``Ours'' denotes boundary-aware KTO.}
    \label{fig:hallucination_rate}
\end{figure}

Table~\ref{tab:alignment_results} and Figure~\ref{fig:hallucination_rate} show the recommendation performance and hallucination rate at different training stages, respectively. In this experiment, the other CARA modules and experimental settings remain unchanged, so the results mainly reflect the effect of the agent alignment stage. We can see that SFT brings stable improvements over the base model, indicating that structured supervised data helps the agent learn the candidate judgment format and basic task logic. After further applying KTO, the model continues to improve on most metrics, suggesting that preference optimization can improve candidate judgments that remain unstable after SFT.

In addition to recommendation performance, KTO further reduces the agent's hallucination rate. Since the candidate judgment rationales generated by the agent serve as the context for subsequent affective and rational judgments, factually inconsistent content may lead to erroneous evidence propagation. Therefore, the reduction in hallucination rate indicates that KTO improves not only task accuracy but also the grounding ability of the recommendation pipeline. Overall, SFT provides format stability and initial decision-making ability, while boundary-aware KTO further improves factual consistency and judgment reliability on complex samples.

\subsubsection{Boundary Instruction Statistics}

Figure~\ref{fig:hallucination_rate} compares vanilla KTO, which applies preference optimization to all candidate instructions, with the proposed boundary-aware KTO. Vanilla KTO does not consistently reduce hallucinations across the three domains, indicating that indiscriminately using all preference samples may introduce weak or noisy optimization signals. In contrast, boundary-aware KTO achieves lower hallucination rates by restricting optimization to instructions with intermediate empirical solvability. This result suggests that the effectiveness of KTO depends not only on the amount of preference data, but also on the informativeness of the selected instructions.

Table~\ref{tab:kto_data_stats} further shows that the selected boundary instructions account for only a small fraction of all candidate instructions. After SFT, many easy instructions can already be solved consistently and therefore provide limited additional learning signals, whereas overly difficult instructions rarely yield reliable desirable responses and may introduce noisy feedback. Boundary-aware selection instead retains instructions that the model can solve occasionally but not consistently. Compared with vanilla KTO, this strategy reallocates the optimization budget toward learnable yet unstable behaviors. The lower hallucination rates obtained with substantially fewer instructions provide evidence that increasing the density of informative preference signals can be more effective than uniformly optimizing over all available instructions.
\begin{table}[!htbp]
    \centering
    \begin{tabular}{lrrr}
        \toprule
        Dataset & Instructions & Boundary & Records \\
        \midrule
        CDs    & 4,435 & 723 & 2,892 \\
        Office & 2,225 & 433 & 1,732 \\
        Beauty & 1,980 & 360 & 1,440 \\
        \bottomrule
    \end{tabular}
    \caption{Statistics of boundary-aware KTO data construction.}
    \label{tab:kto_data_stats}
\end{table}

\section{Conclusion}

We presented CARA, a cognitively adaptive recommendation agent that formulates recommendation as a structured decision process. CARA combines grounded candidate filtering, affective and rational judgment, confidence-adaptive ranking, and error-driven updates of separate affective and rational memories, enabling continuous preference calibration without modifying the underlying language model.

Experiments on three Amazon Reviews domains demonstrate the
effectiveness of CARA under sparse interaction settings. CARA
outperforms representative baselines on most metrics, while ablation
and post-training analyses verify the contributions of its core modules
and show that boundary-aware KTO improves judgment stability and
reduces unsupported generations. Despite these encouraging results,
the current evaluation is limited to three product domains, a relatively
small user population, and sampled candidate sets. Future work will
evaluate CARA in larger-scale, interactive, and cross-domain settings,
while exploring richer feedback signals and more dynamic memory
updates.




\bibliography{aaai2027}


\end{document}